\documentclass{elsarticle}

\usepackage[utf8]{inputenc}
\usepackage[font={footnotesize,it}]{caption}
\usepackage{amsmath,amssymb}
\usepackage{graphicx}
\usepackage{hyperref}

\def\pagenum{p.~}
\def\ppagenum{pp.~}
\def\V{\hspace{0.25mm}'}

\biboptions{authoryear}

\begin{document}

\begin{frontmatter}

\title{Maxwell and the normal distribution: A colored story of probability, independence, and tendency toward equilibrium}
\author{Bal\'azs Gyenis}
\ead{gyepi@hps.elte.hu}
\ead[url]{http://hps.elte.hu/~gyepi}
\address{Institute of Philosophy, Hungarian Academy of Sciences RCH, 4 T\'oth K\'alm\'an Str., Budapest, Hungary}

\begin{abstract}
We investigate Maxwell's attempt to justify the mathematical assumptions behind his 1860 Proposition IV according to which the velocity components of colliding particles follow the normal distribution. Contrary to the commonly held view we find that his molecular collision model plays a crucial role in reaching this conclusion, and that his model assumptions also permit inference to equalization of mean kinetic energies (temperatures), which is what he intended to prove in his discredited and widely ignored Proposition VI. If we take a charitable reading of his own proof of Proposition VI then it was Maxwell, and not Boltzmann, who gave the first proof of a tendency towards equilibrium, a sort of H-theorem. We also call attention to a potential conflation of notions of probabilistic and value independence in relevant prior works of his contemporaries and of his own, and argue that this conflation might have impacted his adoption of the suspect independence assumption of Proposition IV. 
\end{abstract}


\end{frontmatter}

\section{Introduction}

James Clerk Maxwell's early work on the kinetic theory of gases was a major step-stone in the introduction of probabilistic methods into physics. Proposition IV of Maxwell's \citeyearpar{Maxwell1860a} {\it Illustrations of the Dynamical Theory of Gases}, his first derivation of the velocity distribution law, is frequently cited as ``one of the most important passages in physics'' \citep[\pagenum 34]{Truesdell1975} and as such had been thoroughly investigated in the works of S. G. Brush \citeyearpar{Brush1958,Brush1971,Brush1976,Brush1983}, M. C. Dias \citeyearpar{Dias1994}, C. W. F. Everitt \citeyearpar{Brush-Everitt-Garber1986b}, E. Garber \citeyearpar{Garber1970}, C. C. Gillispie \citeyearpar{Gillispie1963}, P. M. Heimann \citeyearpar{Heimann1970}, T. M. Porter \citeyearpar{Porter1981}, C. Truesdell \citeyearpar{Truesdell1975}, J. Uffink \citeyearpar{Uffink2007} and other historians of science. Proposition IV shows that, given certain assumptions, the three velocity components of molecules of a box of gas follow the normal distribution and the speed (the magnitude of velocity) follows what nowadays is called the Maxwell-Boltzmann distribution.

This paper provides a new conceptual and historical analysis of Maxwell's derivation. We make four contributions to the literature. First, the paper sheds new light on the logical structure of Proposition IV and gives the first detailed analysis of how Maxwell attempted to physically justify the three main mathematical assumptions on which it rests. This allows us to show, second, that contrary to the common historical wisdom molecular collisions did play an essential role in establishing the conclusion of Proposition IV: one of its three mathematical assumptions requires a prior lemma showing that collisions among particles of the same mass bring about an equilibrium velocity distribution, and we argue that Maxwell indeed made an attempt to lend credence to such an approach to equilibrium on the basis of his Propositions I--III. 

To substantiate this claim we take a look at Maxwell's remarks preceding Proposition IV as well as at his attempt to arrive at a proof of tendency towards equilibrium in his discredited and widely ignored Proposition VI. We present a surprisingly simple proof of tendency towards equilibrium that rests on Propositions I--III, both in the relevant special case when all particles have the same mass and in the general case when masses differ. Although this proof of equalization of temperatures is interesting in its own right and is not known by experts, we argue that it is not novel since its general case is nothing but a charitable reconstruction of Maxwell's Proposition VI. Since the charitable reading of Proposition VI provides a proof of tendency towards equilibrium in the general case, and since the special case of this proof is really simple, it is reasonable to assume that Maxwell's brief remarks preceding Proposition IV are at least partially motivated by this special case. Additionally, if the charitable reading is tenable then Maxwell preceded Ludwig Boltzmann by at least 6 years in providing a mechanical argument for tendency towards equilibrium, and this priority of Maxwell should be more widely recognized.

Fourth, the paper contributes to the scholarly discussion of the crucial and arguably unjustified probabilistic independence assumption of Proposition IV. The conflation of different interpretations of probability have already been pointed out in the literature as a potential source of Maxwell's mistake in accepting this independence assumption; besides furthering this analysis we also provide a novel interpretation according to which Maxwell, instead of conflating different interpretations of probability, might have conflated different notions of independence. We distinguish two notions -- probabilistic independence and value (or parameter) independence --, argue that their difference was not clear in the relevant works of Clausius and Herschel, and point out that in fact Herschel mistakenly emphasized that it is the satisfaction of value independence that is crucial for the proof that Maxwell allegedly have adapted as his Proposition IV. Thus if Maxwell indeed adapted the proof from Herschel's review article then he also likely to have accepted Herschel's assessment of the assumption on which ``the whole force of the proof turns.'' This in turn could explain why Maxwell was content with the physical justification of the independence assumption since value independence of the velocity components would have appeared immediately clear. Although there seems to be no conclusive evidence deciding which of these two conflations were committed by Maxwell (maybe both), we point out that in his prior and contemporaneous work Maxwell means value independence when he is discussing ``independence''. 

\begin{table}
	\begin{center}
		\begin{tabular}{|l|l|l|}
			\hline
				{\it Illustrations} locus					&	Content												& 	Section(s)																					\\ \hline \hline
				Proposition I--III							&	Analysis of particle collisions; 				&	\ref{sect_propi-iii}, \ref{sect_twonotionsofindependence}, \ref{sect_justification}															\\
															&	{\em Condition M} appears as a hidden			&																										\\
															&	assumption of Proposition II.					&																										\\ \hline
				{\em `If a great many equal} 				&	Claim that collisions would lead 				&	\ref{sect_propvi}, \ref{sect_justification}																\\
				{\em  spherical particles}					&	to	a stable kinetic energy (tempe- 					&																											\\
				{\em were in motion [...]'}					&	rature) and velocity distribution. 						&																									\\ \hline
				Proposition IV								&	Derivation of velocity distribution.			&	\ref{sect_proposition_iv}, \ref{sect_historicaltreatment}, \ref{sect_twonotionsofindependence}, \ref{sect_justification}	\\ \hline
				Proposition V								&	Derivation of relative velocity 				&	(\ref{sect_proposition_iv})																			\\ 
															&	distribution.									&																										\\ \hline
				Proposition VI								&	Collisions lead to an equalization	 			&	\ref{sect_propvi}, \ref{sect_justification}																\\
															&	of temperatures of a mix of gases.				&																										\\ \hline
		\end{tabular}
		\caption{Linear structure of the beginning of the {\em Illustrations of the Dynamical Theory of Gases (\citeyearpar{Maxwell1860a})} with the sections where they are analyzed in this paper.}
	\end{center}	
\end{table}

The paper is organized as follows. Section \ref{sect_proposition_iv} starts with a reconstruction of Maxwell's proof of Proposition IV and identifies the three mathematical assumptions (of existence, independence, and symmetry) that are required for the proof to work. We work backwards from here, first historically, then conceptually. We recall Maxwell's own wording of his proof of Proposition IV and contrast it with a proof of Herschel that has been identified by historians as its likely source. Section \ref{sect_historicaltreatment} also briefly reviews the existing historical literature regarding Maxwell's attempt to justify the mathematical assumptions of Proposition IV. Section \ref{sect_propi-iii} reviews the particle collision model of Maxwell's Proposition I--III and identifies a crucial assumption (Condition M) that is needed for Proposition II to work. Section \ref{sect_propvi} shows how Proposition I--III and Condition M can be used to argue for the equalization of mean kinetic energies and contrasts the argument with Maxwell's Proposition VI. Section \ref{sect_twonotionsofindependence} distinguishes between two independence notions and calls attention to the conflation of these two notions in the works of Herschel, Clausius, and Maxwell. Section \ref{sect_justification} puts the pieces back together by analyzing the extent to which the three aforementioned mathematical assumptions of Proposition IV -- that of existence (Section \ref{subsect_existence}), independence (Section \ref{subsect_independence}), and symmetry (Section \ref{subsect_rotationalsymmetry}) -- could be physically justified by Maxwell's collision model.

\section{Proposition IV -- Maxwell's derivation}\label{sect_proposition_iv}

As we know from a 1859 letter to George Gabriel Stokes \citep[\pagenum 277]{Maxwell1859b}, Maxwell's interest in the kinetic theory of gases was aroused by the papers of Rudolf Clausius. Clausius was mainly interested in explaining heat in terms of molecular motion, and in his 1857 article he used an elastic sphere model to establish a connection between the average kinetic energy and the pressure of the gas. This treatment met the criticism of a Dutch meteorologist, C. H. D. Buys--Ballot \citeyearpar{Buys-Ballot1858}:  if, as calculations suggest, the molecules of a gas move at speeds of several hundreds meters per second, odors released in one corner of a room should almost instantly be noticed in the other corner. To answer the objection, Clausius \citeyearpar{Clausius1859} attempted to show that repeated collisions prevent molecules from traveling for great distances in straight lines. These considerations led him to introduce the notion of the mean free path (for further historical details see \citet{Brush-Everitt-Garber1986b}). Maxwell developed the approach further in his {\em Illustrations of the Dynamical Theory of Gases} \citep{Maxwell1860a}. 

The main goal of Herapath, Waterston, Kr\"onig and Clausius was to establish the kinetic theory as a bridge between thermodynamics and atomic theory. However, the focus of the {\em Illustrations} is on problems of viscosity, diffusion and heat conduction: Maxwell attempted to treat these as special cases of a general process in which momentum or energy is transported by molecular motion. To investigate these transport properties he relied on results he derived about the velocity distribution of a gas in his Proposition IV.

Proposition IV aims to show that in a box of gas the number of particles whose velocity component in a particular direction lies between $v_x$ and $v_x+dv_x$ is proportional to $e^{-\frac{v_x^2}{a^2}}$ and that the number of particles whose speed  lies between $v$ and $v+dv$ is proportional to $v^2 e^{-\frac{v^2}{a^2}}$, where $a$ is a parameter which gets determined later. In essence the proof relies on the fact that the only solution of the functional equation
\begin{equation}\label{functionalequation}
f^2(0) \cdot f(|\vec{v}|) = f(v_x) \cdot f(v_y) \cdot f(v_z)
\end{equation}
is the Gaussian
\begin{equation}\label{gaussian}
f(v_x) = C \cdot e^{A \cdot v_x^2}.
\end{equation}
Equation (\ref{functionalequation}) arises from three assumptions: that
\begin{itemize}
	\item[(A1)] a stationary velocity distribution exists;
	\item[(A2)] the components of velocity in an orthogonal coordinate system are independent; 
	\item[(A3)] the velocity distribution only depends on the magnitude of the velocity. 
\end{itemize}
If we denote the stationary velocity distribution of (A1) with $\mathbf{f}(\vec{v})$ and if by ``independence'' we mean probabilistic independence (treating the velocity components as random variables) then (A2) translates to $\mathbf{f}(\vec{v}) = \mathbf{f}_x(v_x) \mathbf{f}_y(v_y) \mathbf{f}_z(v_z)$ and (A3) translates to rotational symmetry of $\mathbf{f}$, namely that there exists a function $g$ for which $\mathbf{f}(\vec{v}) = g(|\vec{v}|)$. Noting that $\mathbf{f}_x, \mathbf{f}_y$, and $\mathbf{f}_z$ are distributions it is easy to show that $f \doteq \mathbf{f}_x = \mathbf{f}_y = \mathbf{f}_z$ and that $g(.) = f^2(0) \cdot f(.)$, leading to equation (\ref{functionalequation}) . Let us quote Maxwell's own proof (with a minor change of notation). 
\begin{quote}
Prop. IV. To find the average number of particles whose velocities lie between given limits, after a great number of collisions among a great number of equal particles.

Let $N$ be the whole number of particles. Let $v_x$, $v_y$, $v_z$ be the components of the velocity of each particle in three rectangular directions, and let the number of particles for which $v_x$ lies between $v_x$ and $v_x+dv_x$, be $N f(v_x) dv_x$, where $f(v_x)$ is a function of $v_x$ to be determined.

The number of particles for which $v_y$ lies between $v_y$ and $v_y+dv_y$ will be $N f(v_y) dv_y$; and the number for which $v_z$ lies between $v_z$ and $v_z + dv_z$ will be $N f(v_z) dv_z$, where $f$ always stands for the same function.

Now the existence of the velocity $v_x$ does not in any way affect that of the velocities $v_y$ or $v_z$, since these are all at right angles to each other and independent, so that the number of particles whose velocity lies between $v_x$ and $v_x+dv_x$, and also between $v_y$ and $v_y+dv_y$, and also between $v_z$ and $v_z+dv_z$, is
\begin{equation*}
N f(v_x) f(v_y) f(v_z) dv_x dv_y dv_z.
\end{equation*}
If we suppose the $N$ particles to start from the origin at the same instant, then this will be the number in the element of volume $(dv_x dv_y dv_z)$ after unit of time, and the number referred to unit of volume will be 
\begin{equation*}
N f(v_x) f(v_y) f(v_z).
\end{equation*}
But the directions of the coordinates are perfectly arbitrary, and therefore this number must depend on the distance from the origin alone, that is 
\begin{equation*}
f(v_x) f(v_y) f(v_z) = \phi (v_x^2 + v_y^2 + v_z^2).
\end{equation*}
Solving this functional equation, we find 
\begin{equation*}
f(v_x)=C e^{A v_x^2}, \quad \phi(v^2) = C^3 e^{A v^2}.
\end{equation*}
\citep[\ppagenum 289-290]{Maxwell1860a}
\end{quote}
By applying the fact that the total number of particles is finite and fixing the constants, $f(v_x)$ turns out to be the normal distribution, $\frac{1}{a \sqrt{\pi}} e^{-\frac{v_x^2}{a^2}}$. Thus, Maxwell concludes, the number of particles whose velocity in a given direction lies between $v_x$ and $v_x+dv_x$ is
\begin{equation}\label{xesdxkozott}
dN_{v_x}^{v_x+dv_x} = N \frac{1}{a \sqrt{\pi}} e^{-\frac{v_x^2}{a^2}} dv_x
\end{equation}
and the number whose speed lies between $v$ and $v+dv$ is
\begin{equation}\label{vesdvkozott}
dN_v^{v+dv} = N \frac{4}{a^3 \sqrt{\pi}} v^2 e^{-\frac{v^2}{a^2}} dv.
\end{equation}

According to Proposition V we can obtain similar results for the distribution of relative velocities of pairs of randomly chosen particles: the components of relative velocities also follow the normal distribution. Proposition VI intends to show that collisions lead to the equalization of the mean kinetic energies of systems of particles of different masses. From these results, mean values of various functions of velocities used in determining the viscosity and diffusion coefficients could be expressed in terms of $a$, the standard deviation of the velocities. But it's not just the derivation of mean values that had significance, for many physical properties of gases depend on the actual distribution of molecular velocity. Although the qualitative idea that even at constant temperature different molecules have different velocities was often assumed in verbal formulations -- e.g.\ in Clausius' \citeyearpar[\ppagenum 113-116]{Clausius1857} explanation of evaporation -- equations (\ref{xesdxkozott}) and (\ref{vesdvkozott}) are the first quantitative, experimentally verifiable results of the kinetic theory of heat.

\section{Proposition IV -- historical treatment}\label{sect_historicaltreatment}

On the basis of indirect evidence, several authors \citep{Gillispie1963,Brush-Everitt-Garber1986b} have suggested that Maxwell simply adapted the derivation of Proposition IV from a work in statistics, most probably relying on a July 1850 {\em Edinburgh Review} article by Sir William Herschel on Quetelet's collection of essays on probability \citep{Herschel1850}. The review was also reprinted in Herschel's {\em Essays} in 1857, and we know from an early 1858 letter \citep[\ppagenum 301-302]{Campbell-Garnett1882} that Maxwell read and ``liked'' these essays.

In this 57 pages long, typical Victorian-style review, Herschel first gives his own analysis of the subject and praises Quetelet's work on the derivation of the law of least squares. But, to make the subject more palpable to non-mathematical readers, Herschel comes up with a new, ``simple, general, and perfectly elementary proof of the principle in question," based on the following three assumptions:
\begin{quote}
We set out from three postulates. 1st, that the probability of a compound event, or of the concurrence of two or more independent simple events, is the product of the probabilities of its constituents considered singly; 2nd, that there exists a relation or numerical law of connexion (at present unknown) between the amount of error committed in any numerical determination and the probability of committing it, such that the greater the error the less its probability, according to some regular LAW of progression, {\em which must necessarily be general and apply alike to all cases, since the causes of error are supposed alike unknown in all;} and it is on this ignorance, and not upon any peculiarity on cases, that the idea of probability in the abstract is founded; 3dly, that errors are equally probable if equal in numerical amount, whether in excess, or in defect of, or in any way beside the truth. This latter postulate necessitates our assuming the function of probability to be what is called in mathematical language {\em an even function,} or a function of the square of the error, so as to be alike for positive and negative values; and the postulate itself is nothing more than the expression of our state of {\em complete} ignorance of the causes of error, and their mode of action. \citep[\ppagenum 19-20]{Herschel1850}
\end{quote}
In the course of the derivation Herschel adds that the independence assumption applies to deviations in a rectangular direction: ``the observed oblique deviation is equivalent to the two rectangular ones, supposed concurrent, and which are essentially independent from one another" \citep[\pagenum 20]{Herschel1850}. 
Together with this later addition Herschel's three main assumptions -- existence of a distribution of deviations, probabilistic independence of the deviations in a rectangular direction, and spherical symmetry of the distribution -- are similar to those of Maxwell. Herschel's derivation of the functional form, albeit in a verbalized form, proceeds in the same way as Maxwell's, the main difference being that Herschel considers two rather than three dimensions and that he shows no interest in calculating the distribution of the magnitude of the errors.

Even if Maxwell adopted Herschel's derivation it is natural to ask whether Maxwell tried to justify the mathematical assumptions of Proposition IV. According to Maxwell the conclusion of Proposition IV is intended to apply ``after a great number of collisions among a great number of equal particles'' takes place. Proposition IV of the {\em Illustrations} is preceded by three other propositions that describe the nature and effects of particle collisions. Yet we find that the historical literature pays very little to no attention to these prior findings of Maxwell; we typically find half-sentence-remarks about the role Propositions I--III played in the argumentation, with as many different half-sentence-remarks as there are historians making them. 

The only pertaining remark \citet[\ppagenum 98-99]{Porter1981} -- who explicitly deals with the history of statistics and probability theory that lies in the background of Proposition IV -- makes is that Maxwell's ``kinetic gas model [...] was subject to a nearly infinite number of minute causes of deviation, and it displayed perfect statistical regularity,'' and this somehow ``led Maxwell to the insight that the distribution of molecular velocities probably conformed to [the error] curve.'' It is rather unclear from this remark how Maxwell is supposed to have been led to his insight or even just what the insight of Maxwell was; understood strictly Porter's description is misleading since Maxwell did not make a probabilistic statement about the conformity of the actual velocity distribution to the normal distribution. 

\citet[\pagenum 34]{Truesdell1975} mentions that Maxwell's work ``begins [...] with consideration of the impact of two spheres" but he doesn't attribute any importance to this part of Maxwell's paper. 

Other historians generally assume that Maxwell's analysis of molecular collisions has something to do with underpinning Proposition IV, but we find no clear consensus about the motivations. The longest treatment, referred to by many subsequent historical articles, comes from \citet[\pagenum 243-244]{Brush1958} (repeated almost verbatim in \citet[\ppagenum 185-186]{Brush1976}) whose length still permits it to be quoted in full: 
\begin{quote}
[Maxwell] began by pointing out that, if two elastic spheres collide, all directions of rebound are equally likely. He apparently believed that this fact would ensure, not only that all directions of motion were equally probable in the gas, but also that the probability distribution for each component of the velocity was independent of the values of other components.
\end{quote}
In other words Brush suggests that Maxwell's analysis of particle collisions was aimed at justifying the probabilistic independence assumption (A2) and the rotational symmetry assumption (A3) but he provides no details about how Maxwell's justification is supposed to have proceeded. It is also not clear what Brush means by ``pointing out''; as we will see later in this respect \citet[\pagenum 194]{Heimann1970} is slightly more accurate when he says in his half sentence remark that ``Maxwell assumed that all motions [of the particles] were equally probable'' (although, as we will see, Maxwell's {\em assumptions} only {\em implied} that after rebound all {\em directions} of motion are equally probable). 

According to \citet[\pagenum 7]{Brush-Everitt-Garber1986b} Maxwell claimed that ``when a moving sphere collides with with another fixed sphere randomly placed in its path, the direction in which it rebounds is distributed with equal probability over each element of solid angle'' (this is incorrect since Maxwell did not assume any of the particles to be fixed), and they further state -- without providing any further details or evidence -- that Maxwell used this claim to argue that ``the distribution law has an explicit functional form''.  

\citet[\pagenum 944]{Uffink2007} mentions en passant that Maxwell realized that the ``velocities [of the particles] will suffer incessant change due to mutual collisions'' but he implies that Maxwell's reference to collisions is, with Maxwell's own remark made 20 years later about certain derivations in the works of others, ``rather for the sake of enabling the reader to form a mental image of the material system than as a condition for the demonstration'' (ibid.\ p.\ 948). Uffink also emphasizes that the assumptions of Maxwell's Proposition IV, despite appearances and in contrast to Maxwell's subsequent proof of 1867, ``do not refer to collisions at all'' (ibid. p. 948; see also his contrast with Maxwell's 1867 proof on ibid.\ p.\ 950). 

\citet{Dias1994} is the only exception in giving an extensive treatment of Maxwell's collision model but the focus of the paper is on exploring the connections between Maxwell's Proposition II and later debates of the 1890s on the Stosszahlansatz; Dias does not address what role, if any, Maxwell's collision model played in justifying the assumptions of Proposition IV.

The scant attention paid to the justification question may stem from historians unanimously following Brush's \citeyear{Brush1958} reconstruction of Proposition IV as a proof resting on {\em two} assumptions, that of probabilistic independence (A2) and of rotational symmetry (A3), and they are at best only tacit about the need for a third assumption (A1) that a stationary velocity distribution exists (see i.e. \citet[\ppagenum 185-186, \ppagenum 342-343]{Brush1976}, \citet[\pagenum 194]{Heimann1970}, \citet[\pagenum 98]{Porter1981}, \citet[945]{Uffink2007}). They do so despite acknowledging the analogy with Herschel's derivation pointed out by Gillispie, five years after Brush's \citeyear{Brush1958} reconstruction, in \citeyear{Gillispie1963}. As we have seen Hershel explicitly emphasized a third assumption, that there must ``exists a relation or numerical [...] regular LAW of progression, {\em which must necessarily be general and apply alike to all cases}'' \citep[\ppagenum 19-20; emphasis in original]{Herschel1850}, which adapted to the kinetic theory translates as the assumption of the existence of a stationary velocity distribution, our (A1). By only explicitly acknowledging two assumptions instead of three the natural guess might have been that what Maxwell tried to do with his collision model was to physically justify one or both of these two assumptions. Since without a detailed analysis of Maxwell's collision model only superficial analogies are apparent between Maxwell's prior results and assumptions (A2)--(A3) these prior results of Maxwell were supposedly aimed at justifying, and since Maxwell himself later abandoned the independence assumption (A2) describing it as being ``precarious'' \citep[\pagenum 437]{Maxwell1867}, it could have been tempting to glance over the opening part of Maxwell's paper as being non-consequential to the argument presented in Proposition IV. 

We are going to argue that the most plausible role the collision model played in Maxwell's thinking is in justifying assumption (A1), although our analysis of Maxwell's collision model also sheds light on the extent to which his prior results are capable to support assumptions (A2) and (A3) beyond the superficial analogies. To do this we start with a brief reconstruction of his Propositions I--III.

\section{Propositions I--III and Condition M -- particle collisions}\label{sect_propi-iii}

\begin{figure}
	\begin{center}
		\includegraphics[height=4.5cm]{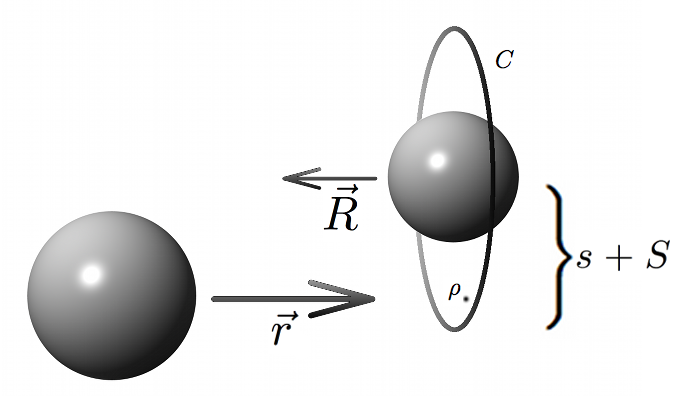}
		\caption{$\vec{r}, \vec{R}$: velocities relative to the center of mass. $C$: circle of radii $s+S$ centered on the second particle in the plane perpendicular to $\vec{r}$. $\rho$: point where the line of motion of the first particle intersects circle $C$.}\label{propiifigure}
	\end{center}
\end{figure}

In the beginning of the first chapter, ``On the Motion and Collision of Perfectly Elastic Spheres,'' Maxwell presents us with three Propositions concerning the interaction of particles. Proposition I and Proposition III are straightforward exercises in Newtonian mechanics that establish the motion of two particles -- modeled by perfectly elastic spheres of non-zero radii $s$, $S$ and of mass $m$, $M$ -- after they collide. It is assumed that the particles do not interact aside such pairwise collisions. In modern notation Maxwell calculates the encounter operator $(\vec{v}, \vec{V}, \rho) \mapsto  (\vec{v}\V, \vec{V}')$ where $\vec{v}, \vec{V}$ are the velocities of the particles before the impact, $\vec{v}\V, \vec{V}'$ are the velocities after impact, and $\rho$ is a further parameter describing the impact in the following way. Let $\vec{r} = \vec{v} - \vec{v}_{CM}$ be the velocity of the first particle relative to the velocity of the center of mass $\vec{v}_{CM} = \frac{1}{m+M} (m \vec{v} + M \vec{V})$ before impact and let $C$ be a circle of radius $s+S$ that lies in the plane perpendicular to $\vec{r}$, $C$ being centered on the center of the second particle. For a collision to take place the line of motion of the first particle needs to intersect $C$;  $\rho$ is the position of this intersection within $C$. (See Figure \ref{propiifigure}.)

Due to conservation of momentum the collision does not affect the velocity of the center of mass: $\vec{v}\V_{CM} = \vec{v}_{CM}$. Maxwell shows that a perfectly elastic collision alters the direction of the relative velocity but it keeps its magnitude intact:
\begin{equation}\label{eq_r_is_unchanging}
\vec{r}\V  = r  \cdot \vec{u}(\rho),
\end{equation}
and therefore the velocity after impact can be expressed as
\begin{equation}\label{velocityafterimpact}
\vec{v}\V  = \vec{v}_{CM} + r \cdot \vec{u}(\rho),
\end{equation}
where $r = |\vec{r}|$ and $\vec{u}(\rho)$ is a direction of unit length depending on the position $\rho$ within the circle $C$. (Mutatis mutandis for the  $\vec{R} = \vec{V} - \vec{v}_{CM}$ velocity of the second particle relative to the center of mass.)

Maxwell also claims to have proven that whenever a collision takes place all directions of rebound are equally likely in the center of mass frame. What his Proposition II actually shows is that if the assumption that 
\begin{description}
	\item[Condition M] the impact parameter $\rho$ is distributed uniformly within the circle $C$
\end{description}
holds then
\begin{description}
	\item[Condition A] the $\vec{u}(\rho)$ directions of rebound relative to the center of mass are distributed uniformly on the surface of the unit sphere. 
\end{description}
The name Condition A is due to S. H. Burbury \citeyearpar{Burbury1894a} who invoked it in a debate on the pages of {\em Nature} with G. H. Bryan about Boltzmann's H-theorem. Condition M is the terminology we introduce here to distinguish Condition A from the assumption Maxwell actually made during the proof of Proposition II. Since in Maxwell's particle model Condition M and Condition A imply each other they tend to be identified (for such an exchange of conditions see i.e. \citet[\pagenum 347]{Dias1994}). Maxwell himself neither emphasizes nor attempts to justify Condition M; he merely asserts that ``within this circle [$C$] every position [$\rho$] is equally probable'' \citep[\pagenum 288]{Maxwell1860a}, and in the coming years he keeps referring to the conclusion of Proposition II as if it were unconditional (see e.g. \citet[\pagenum 339]{Maxwell1863}; for a further analysis of Condition A and the {\em Nature} debate see \citet{Dias1994}).

The conclusion drawn from Propositions I--III is that the $\vec{v}\V$ velocity of a particle after impact consists of two components: the first is the $\vec{v}_{CM}$ velocity of the center of mass, which is determined by the $\vec{v}$ and $\vec{V}$ incoming velocities, and the second is the $\vec{r}\V$ velocity relative to the center of mass, whose magnitude is determined by the incoming velocities, but whose direction is not determined by the incoming velocities, but may with equal probability be in any direction.

\section{Proposition VI -- tendency towards equilibrium}\label{sect_propvi}

After being done with Propositions I--III, but before turning his attention to Proposition IV, Maxwell makes the following remark about the effect of many collisions:
\begin{quote}
If a great many equal spherical particles were in motion in a perfectly elastic vessel, collisions would take place among the particles, and their velocities would be altered at every collision; so that after a certain time the {\em vis viva} will be divided among the particles according to some regular law, the average number of particles whose velocity lies between certain limits being ascertainable, though the velocity of each particle changes at every collision. \citep[\pagenum 289]{Maxwell1860a}
\end{quote}
Thus Maxwell claims that after sufficient number of collisions the kinetic energies become distributed along a non-changing distribution. This implies that some sort of process that brings about this distribution of kinetic energies takes place, apparently due to the collisions these particles go through.

On the surface of it Maxwell's remark strikes as an expression of hope rather than an argument. Is there any evidence that Maxwell had good reasons to hold that Propositions I--III are capable of supporting a claim about tendency towards equilibrium? The answer is affirmative, and this becomes clear if we jump ahead to Proposition VI in which Maxwell argues that when two gases mix their temperatures equalizes. This argument is unknown in the historical and in the physics literature and since besides being relevant to the question of justification of an assumption of Proposition IV it is also interesting and historically significant on its own right, we now briefly reconstruct it.

\subsection{The reconstructed argument for tendency towards equilibrium}\label{subsect_propvi_reconstructed}

When two particles, modeled as perfectly elastic spheres of mass $m$ and $M$, collide the difference between their kinetic energies after the collision can be readily expressed as a function of the difference between their kinetic energies before the collision:
\begin{equation}\label{eq_differenceofkineticenergiesafterrebound}
\left( \frac{m v\V^2}{2} - \frac{M V\V^2}{2} \right) = C_0 \cdot \left( \frac{m v^2}{2} - \frac{M V^2}{2} \right) + C_1 \cdot \cos \alpha + C_2 \cdot \cos \gamma, 
\end{equation}
where $\alpha$ is the angle enclosed by the incoming velocities whose magnitude is $v$ and $V$ and $\gamma$ is the angle enclosed by the velocity of the center of mass $\vec{v}_{CM}$ with the relative rebound velocity $\vec{r}\V$ (see Figure \ref{propvifigure}; for calculation of the coefficients see the Appendix). Propositions I--III together with Condition M entail that $\vec{r}\V$ may point to all directions with equal probability, which means there is an equal chance that the angle of rebound $\gamma$ falls between $\bar{\gamma}$ and $\bar{\gamma} + d\bar{\gamma}$ or between $180^o - \bar{\gamma}$ and $180^o - \bar{\gamma} - d\bar{\gamma}$.

\begin{figure}
	\begin{center}
		\includegraphics[height=4.5cm]{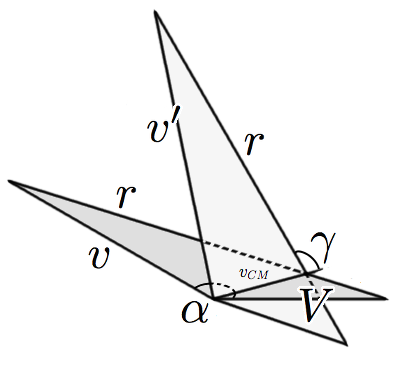}
		\caption{$\vec{v}, \vec{V}$: incoming velocities; $\vec{v}_{CM}$: velocity of center of mass; $\vec{r} = \vec{v} - \vec{v}_{CM}$: velocity of the first particle relative to the center of mass; $\vec{v}\V$: velocity of the first particle after rebound; $\alpha = \angle (\vec{v},\vec{V})$: angle between incoming velocities; $\gamma = \angle (\vec{r}\V,\vec{v}_{CM})$: angle of rebound.}\label{propvifigure}
	\end{center}
\end{figure}

Consider now collisions between many particles during a time interval $\Delta t$, and let us write $(k,l) \in I_{v,V}^{\alpha}$ whenever $k$ is a particle of mass $m$, $l$ is a particle of mass $M$, and $k$ and $l$ collide during $\Delta t$ with respective incoming speeds $v$, $V$ and angle of income $\alpha$. We assume that $\Delta t$ is long enough for $|I_{v,V}^{\alpha}|$ to be large but short enough to allow for at most one collision for each particle. Let us sum up equations (\ref{eq_differenceofkineticenergiesafterrebound}) for all collision pairs in $I_{v,V}^{\alpha}$ by assuming that Condition M holds for all these particle collisions independently. Since then all directions of rebound occur with equal frequency and since $C_2$ only depends on $m, M, v, V$, and $\alpha$ which are now all fixed, the $C_2 \cdot \cos \gamma$ terms approximately cancel and we get
\begin{equation}
\sum_{(k,l) \in I_{v,V}^{\alpha}} \left( \frac{m v_k'^2}{2} - \frac{M V_l'^2}{2} \right) \approx  C_0 \cdot |I_{v,V}^{\alpha}| \cdot \left( \frac{m v^2}{2} - \frac{M V^2}{2} \right) + C_1 \cdot |I_{v,V}^{\alpha}| \cdot  \cos \alpha .
\end{equation}

Let us further assume that all directions of incoming velocities for the colliding particles occur with approximately equal frequency -- an assumption that Maxwell took for granted in his later derivations, i.e.\ in \citet{Maxwell1867} -- and that the number of collisions is independent of the enclosed angle of income (that $|I_{v,V}^{\alpha}|$ is the same for all $\alpha$, or at least that $|I_{v,V}^{\alpha}| = |I_{v,V}^{-\alpha}|$ for all $\alpha$ given a fixed $v$ and $V$) and sum up for all possible angles of income. Since $C_1$ only depends on $m, M, v$, and $V$ which are now all fixed, the $C_1 \cdot |I_{v,V}^{\alpha}| \cdot \cos \alpha$ terms approximately cancel and we get 
\begin{equation}
\sum_{\alpha} \sum_{(k,l) \in I_{v,V}^{\alpha}} \left( \frac{m v_k'^2}{2} - \frac{M V_l'^2}{2} \right) \approx C_0 \cdot \sum_{\alpha} |I_{v,V}^{\alpha}| \cdot \left( \frac{m v^2}{2} - \frac{M V^2}{2} \right) .
\end{equation}
By further summing up for all possible magnitudes of incoming velocities $v, V$ on both sides of the equation we account for all collisions between particles of mass $m$ and $M$ during $\Delta t$ exactly once, so a division with the total number of such collisions and substituting for $C_0$ yields
\begin{equation}\label{eq_kineticenergydifferencedecreases}
	\widehat{\frac{m v'^2}{2}} - \widehat{\frac{M V'^2}{2}} \approx \left(\frac{m - M}{m+M}\right)^2 \cdot \left( \widehat{\frac{m v^2}{2}} - \widehat{\frac{M V^2}{2}} \right),
\end{equation}
meaning that the difference between the mean kinetic energies of the colliding particles decrease after the collisions! 

Any change in the difference between the mean kinetic energies (between the temperatures) of {\em all} $m$-particles and of {\em all} $M$-particles during $\Delta t$ is due to collisions between particles of different mass (expressed by equation (\ref{eq_kineticenergydifferencedecreases})) since the total kinetic energy of particles which do not collide and the total kinetic energy of those particles that collide with another particle of the same mass does not change during $\Delta t$. If the particles that undergo collision during $\Delta t$ are randomly drawn from the set of all particles then the larger the initial temperature difference between $m$-particles and $M$-particles the larger the chance that the drawn $I_{v,V}^{\alpha}$ samples of colliding pairs also exhibit a mean kinetic energy difference of the same sign. If the sign is the same then since due to (\ref{eq_kineticenergydifferencedecreases}) the mean kinetic energy difference between the colliding pairs decreases during $\Delta t$ it follows that the temperature difference of all $m$-particles and $M$-particles also decreases. (If the sign is different then the temperature difference between the gases {\em increases}, underlining that the present argument for tendency towards equilibrium is statistical.) By repeating the argument for successive $\Delta t$ time intervals it follows that the difference between temperatures of $m$-particles and $M$-particles tends to vanish. When the temperature difference vanishes or becomes small the collisions continue to produce small temperature fluctuations (due to the mean kinetic energy differences that may still occur in the random sample of colliding particle pairs).

Note that the proof becomes extremely simple when we mix gases of different temperature whose particles have the same mass. In this case equation (\ref{eq_differenceofkineticenergiesafterrebound}) simplifies to
\begin{equation}\label{eq_differenceofkineticenergiesafterreboundsimplified}
\left( \frac{m v\V^2}{2} - \frac{m V\V^2}{2} \right) = C_2 \cdot \cos \gamma
\end{equation}
and one only needs to invoke the first cancellation argument provided by Condition M to conclude that the temperature difference decreases as collisions take place. What we thus have is a simple, elegant, and elementary mechanical argument for a statistical tendency towards equilibrium that is based on the assumptions that 
\begin{itemize}
	\item[(i)] particles can be modeled as perfectly elastic spheres, 
	\item[(ii)] the number, size, speed etc. of these particles is such that the time evolution can be split into successive time intervals during which each particle has an approximately equal independent chance to participate in at most one, pairwise collision, and 
	\item[(iii)] in this interval all directions of rebound occur with approximately the same relative frequency (that for any pair of incoming velocity and for the colliding particle pairs that have these incoming velocities the relative frequency of the impact parameter is roughly the same for all positions on the collision circle). 
\end{itemize}	
Although we know well from later developments that assumptions (ii) and (iii) are difficult to square with the strict dynamical evolution entailed by (i), its simplicity and its geometrically intuitive probabilistic assumptions could very well allow the argument to fit in an undergraduate course on classical mechanics.

\subsection{Proposition VI in the light of the reconstruction}\label{subsect_propvi_real}

How can it be that this simple argument for tendency towards thermal equilibrium is unknown by historians and physics educators alike? The conclusion that collisions bring about an equalization of temperatures is what Maxwell draws in his Proposition VI after he derives a formula he interprets as (\ref{eq_kineticenergydifferencedecreases}) by invoking his earlier results about collisions. Prima facie, however, Maxwell's proof seems rather baffling. What he seems to have calculated in his Proposition VI is the difference of kinetic energies of particles of mass $m$ and $M$ after they collide in a specially arranged manner where (using the notation of the previous subsection)
\begin{itemize}
	\item[(p1)] $\gamma = 90^o$,
	\item[(p2)] $\alpha = 90^o$,
	\item[(p3)] $v = \widehat{v}$, $V = \widehat{V}$.
\end{itemize}
By substituting assumptions (p1)-(p3) to equation (\ref{eq_differenceofkineticenergiesafterrebound}) one gets
\begin{equation}\label{eq_Maxwellreally}
	\frac{m \widehat{v}'^2}{2} - \frac{M \widehat{V}'^2}{2} = \frac{(m - M)^2}{(m + M)^2} \left( \frac{m \widehat{v}^2}{2} - \frac{M \widehat{V}^2}{2} \right),
\end{equation}
which is the equation that Maxwell's own derivation would have really established. Maxwell curiously assumes that $\widehat{v}\V = \widehat{v\V}$ and he also interprets (\ref{eq_Maxwellreally}) as if he had just derived (\ref{eq_kineticenergydifferencedecreases}) and proceeds to draw the conclusion about the long term equalization of mean kinetic energies on the basis of it.

Thus the calculations of Proposition VI make Maxwell's argument seem to be dependent on a particular arrangement of incoming velocities and rebound angle of a pair of particles, and it seems entirely unclear how he thought that an argument about a particular collision of a pair of particles could allow for a valid inference about averages over all collisions, or how the magnitude of velocity after impact that initially equals the mean velocity could be substituted for the mean of velocities after impact, or how the kinetic energy calculated from a mean velocity could be substituted for a mean kinetic energy. On this basis Maxwell's proof, if mentioned at all, is dismissed in the historical literature as a ``rather lame argument'' 
and the few historians who do devote more than a footnote to it find it ``amazing'' that the otherwise brilliant Maxwell or ``any of his contemporaries who bothered to examine the argument in detail would have accepted it'' \citep[\pagenum 344]{Brush1976}.

In the light of the simple argument for equalization of kinetic energies we outlined in the previous subsection -- an argument that rest purely on Maxwell's Propositions I--III and invokes the same sort of calculations that Maxwell performs during his proof of Proposition VI -- it seems plausible that Maxwell was being too terse instead of lame. It is clear from the geometry of collisions and from the argument we gave above that for equation (\ref{eq_differenceofkineticenergiesafterrebound}) averaging over all directions of rebound yields the same result as making assumption (p1), as a result of which the third term cancels. This then justifies focusing on collisions with perpendicular rebound angle as being sufficiently typical for the purposes of the proof. It is also geometrically clear that averaging over all directions of incoming velocities yield the same result as making assumption (p2), as a result of which the second term cancels. This then would justify focusing on collisions with perpendicular income angle as being sufficiently typical for the purposes of the proof. 

Propositions I--III motivates assumption (p1) but assumption (p2) is in need of justification. Instead of invoking a further assumption that all directions of incoming velocities are equally likely (like we did above, and as Maxwell did in his later works), Maxwell sets the magnitude of the incoming velocities $v$, $V$ to equal their mean values given by Proposition IV (assumption (p3)), and he further assumes (p4) that the magnitude of the relative velocity $|\vec{v} -\vec{V}|$ equals its mean value given by Proposition V.  Since Proposition V shows that the square of the mean relative velocity equals the sum of squares of the mean velocities, and therefore that the two incoming velocities must be perpendicular, (p3) and (p4) entail (p2). Assumptions (p3) and (p4) are also assumptions about typical velocities, but they are not appropriate for the purposes of the proof. Albeit the proof that proceeds directly along (p3) and (p4) is clearly suspect, (p3) and (p4) might have been intended as mere short-cut for a more detailed argument that rests on an averaging procedure over the incoming velocities, along the lines we sketched above.

Albeit this reconstruction still renders the proof of Proposition VI messy and misleading, it allows us to see the arrangement choices Maxwell makes during his derivation as geometrically intuitive shortcuts to an averaging procedure that rests on an application of the results of his Propositions I--III. If we make the charitable assumption that Maxwell simply didn't properly clean up his proof, as opposed to making assumptions whose fallaciousness should have been obvious, then his reconstructed Proposition VI becomes a convincing demonstration of a tendency towards equilibrium that lives up to the standards of its own time. Although the proof does depend on unjustified probabilistic assumptions such as Condition M, in this regard it does not fare worse than Boltzmann's later attempts (starting with \citet{Boltzmann1868}) to arrive at his celebrated H-theorem, and thus it should reclaim its historical precedence over Boltzmann's demonstrations as being the first good ``bad'' mechanical proof of tendency towards equilibrium. 

The only historical side-remark I'm aware of that also suggests a direct connection between Maxwell's Propositions I--III and Boltzmann's H-theorem is a side remark of \citet[\pagenum 347]{Dias1994} who says that
\begin{quote}
Maxwell (1860) invoked Proposition II to show how collisions randomized the directions of motion of the molecules, bringing about isotropy of pressure, as demanded by the equation of state for perfect gases. In this use, Proposition II becomes a kind of ``proof'' of how equilibrium of pressure (transmission of linear momentum) sets in, and thus is, metaphorically speaking, a sort of H-Theorem.
\end{quote}
Although the connection between Proposition II and the randomization of direction of motions of molecules is clear, I find no evidence for Dias' claim that Maxwell made any use of Proposition II to argue for the onset of the isotropy of pressure (Dias provides none). From Dias' reference to linear momentum I suspect that his remark was based on a misinterpretation of what Proposition VI aimed to show. 

Maxwell never seem to have revisited Proposition VI. Around 1863 he became dissatisfied with Clausius' mean free path method and started to develop a theory valid for any kind of force laws acting between particles, which he came to treat as centers of force rather than colliding spheres with determinate radii. With his new approach serious problems emerge concerning Proposition I and III: as Maxwell's \citeyear{Maxwell1867} calculations show, the motion of particles subject to a never-vanishing repulsive force law in general depends on their incoming relative velocity, so the angle of rebound also depends on this incoming relative velocity. Although in his {\em On the Dynamical Theory of Gases} \citeyearpar{Maxwell1867} he chooses a particular inverse $5$th-power force law for which the angle of reflection is independent of the incoming relative velocity, this choice is poorly motivated and gets abandoned later. His writings after 1867 indicate that he grew skeptical of the possibility of justifying the Second Law on a mechanical basis \citep[\ppagenum 951--952]{Uffink2007}.

\section{Value vs.\ probabilistic independence}\label{sect_twonotionsofindependence}

Before analyzing the extent to which Proposition I--III, Condition M, and the argument for an approach to equilibrium can be used to justify the mathematical assumptions of Proposition IV it is worthwhile to take a look at how Maxwell, Herschel, and Clausius understood what these mathematical assumptions, in particular the independence assumption (A2), amounted to.

Let us start with Herschel's understanding of the independence assumption of his own proof. We note first that there is a small but important difference between Herschel's original \citeyear{Herschel1850} review article and its \citeyear{Herschel1857} reprint, which is the version that Maxwell have likely read. In a footnote added to the \citeyear{Herschel1857} reprint Herschel attempts to clarify what he means by independence of the deviations of errors in different directions and emphasizes the importance of this independence assumption: 
\begin{quote}
	That is, {\em the increase or diminution [of the deviation] in one or [sic!] which may take place without increasing or diminishing the other.} On this, the whole force of the proof turns. \citep[\pagenum 400]{Herschel1857}, \citep[\pagenum 11]{Brush-Everitt-Garber1986b}. 
\end{quote}

Although Maxwell's likely adaptation of Herschel's derivation became common wisdom in the historical literature, so far it has gone unnoticed by historians that in the quoted emphatic footnote Herschel actually mischaracterizes the crucial necessary independence assumption of the derivation. A mathematically correct derivation requires {\em probabilistic independence} of deviations in rectangular directions; according to Herschel, however, the whole force of the proof turns on the assumption that a deviation in one direction does not impose a restriction on the values a deviation in another direction may take, which is not the same as probabilistic independence of the deviations. 

Let us briefly distinguish these notions of independence. Let $\xi$ and $\upsilon$ denote physical quantities (i.e. velocity components) that can take values in the set $X$ and $Y$ respectively. Suppose that we may not know what other quantities serve as inputs that determine the momentary values of $\xi$ and $\upsilon$, but let us assume that we might know about certain constraints that connect the values $\xi$ and $\upsilon$ can take in the following way. Let $d: X \rightarrow {\cal P}(Y)$ assign to each value $x \in X$ the set of values $d(x) \subseteq Y$ that variable $\upsilon$ may take if we condition upon $\xi$ taking the value $\xi = x$. If $d$ is single valued for all $x \in X$ then if $\xi$ takes a value $x$ then $\upsilon$ must take the only possible value in $d(x)$: we may thus say that {\em $\xi$ value determines $\upsilon$}. If $d(x)$ has more than one element for at least one $x \in X$ then {\em $\upsilon$ is partially value independent from $\xi$}: for some value variable $\xi$ may take there are several possibilities that $\upsilon$ may take that are not determined by the value of $\xi$. If $d(x) = Y$ for all $x\in X$ then {\em $\upsilon$ is value independent from $\xi$}: no matter what value variable $\xi$ takes that does not influence the possible values $\upsilon$ may take. Finally when $\xi$ and $\upsilon$ are random variables then {\em $\upsilon$ is probabilistically independent from $\xi$} if for all $x,y$: $Pr(\xi \leq x \wedge \upsilon \leq y) =  Pr(\xi \leq x) \cdot Pr(\upsilon \leq y)$, or, alternatively, when $g_{\xi,\upsilon}(x,y) = g_{\xi}(x)g_{\upsilon}(y)$ for the respective density functions.

These independence notions are not equivalent with each other. Value independence does not imply probabilistic independence: indeed the domains of two value independent variables do not even need to be endowed with a probability measure. Probabilistic independence also does not imply value independence: a value $x$ of $\xi$ may prevent $\upsilon$ taking a value $y$ without ruining probabilistic independence of $\xi$ and $\upsilon$ if the probability of the occurrence of $x$ and $y$ is zero. Value independence is symmetric, and value independent variables are also partially value independent but the converse is not the case.

It is clear both from Herschel's derivation and from Maxwell's proof of Proposition IV that their ``independence'' assumption is meant to entail the factorization of the probability distribution function. If by ``independence'' one means probabilistic independence then factorization does follow and the proof is correct; factorization however does not follow from value independence, which Herschel mistakenly emphasized to be the crucial assumption of his derivation. 

Now even if Maxwell adapted the derivation from Herschel's article it is natural to assume that he had reasons to believe that the required mathematical assumptions can be motivated by physical considerations about gases. Without such physical considerations the application of the mere mathematical proposition to gases would remain unjustified. But Maxwell's attempt at physically justifying independence of the velocity components would have clearly depended on his understanding of the sense of independence he had believed to guarantee the success of the proof. We are going to return to this point in Section \ref{subsect_independence}.

Let us now turn to Clausius' understanding of the meaning of independence. In his summary of Maxwell's Propositions I--III Clausius (\citeyear[\pagenum 424]{Clausius1862}) claims that the velocity of center of mass $\vec{v}_{CM}$  quantifies the extent to which the velocity after impact is ``dependent'' upon the velocities before impact while the velocity relative to the center of mass $\vec{r}\V$ quantifies the extent to which the velocity after impact is ``independent'' from the velocities before impact. It is not clear what Clausius meant here by ``independence'' since what he writes is compatible with several different readings. Given Condition M $\vec{r}\V$ may point in any direction no matter what the incoming velocities $\vec{v}$ and $\vec{V}$ were, so it is value independent since setting values for $\vec{v}$ and $\vec{V}$ does not restrict the range of possible values for the direction of $\vec{r}\V$. In the same vein $\vec{v}\V$ is partially value independent from $\vec{v}$ and $\vec{V}$ since setting values for $\vec{v}$ and $\vec{V}$ does narrow the range of possible values of $\vec{v}\V$ but $\vec{v}\V$ may still take several values, not just one. Clausius could also have meant that the direction of $\vec{r}\V$ is probabilistically independent from $\vec{v}$ and $\vec{V}$: conditioning upon any value for $\vec{v}$ and $\vec{V}$ does not alter the {\em probability} that $\vec{r}\V$ points in a certain direction.

Upon a closer look what Clausius emphasizes in the text and in the accompanying footnote is that the incoming velocities do not determine the direction of the velocities relative to the center of mass after impact and in turn they do not fully determine the velocities after impact either, as these latter may take several (indeed, infinitely many) different possible values for any fixed values of the incoming velocities. These remarks of Clausius are suggestive that he intended to make a point about the partial value independence of the velocities before and after impact, but there is no evidence that Clausius even distinguished between value independence and probabilistic independence.

Finally let us make a brief remark about Maxwell's own usage of the term ``independent''. The term appears once more in the context of a generalization of Proposition IV and Proposition V (ibid.\ p.\ 315) but the appearance does not shed any more clarity on its intended interpretation than the proof of Proposition IV. In all other contexts in which the term ``independent'' appears in the {\em Illustrations} (see p.\ 288, 298, 300) it is clear that Maxwell understands it as referring to value independence. Taking a systematic look at Maxwell's prior and contemporary work on color vision (see esp.\ \citep[\ppagenum 120-121]{Maxwell1855a}) I also find that Maxwell systematically uses the term independent in the sense of value independence.\footnote{Judging by the number of his publications the topic Maxwell most vehemently pursued in the years prior to and contemporaneous with his early research on the kinetic theory was that of color vision. A previous draft of this paper argued that Maxwell's own experimental work on color vision provided him familiarity with statistical reasoning; in particular Maxwell have encountered a use of distributions which is not rooted in error, ignorance, or uncertainty, and hence understanding probabilities as being rooted in the physical phenomena was not entirely new to him when he started working on the kinetic theory. On the basis of his drawn figures it seems likely that this is also true in particular for the normal distribution. The draft addressed the parallels Maxwell draws between color mixing in his three dimensional color space and addition of vectors in three dimensions in mechanics. Finally it also elaborated on the claim that in these prior works Maxwell understands the term independence as value independence. At the request of a referee these sections got relocated in a separate historical note.}

Indeed I find that ambiguous usage of the term ``independent'' is pervasive in and before the 1860s. That different concepts are named by the same term would not in itself be a problem if users of the term were conscious of their non-interchangeability. Unfortunately it seems that this consciousness was missing in the works of those, such as Clausius and Herschel, that most likely have influenced Maxwell's thinking, and there is no evidence that Maxwell was aware of such distinction in and before his 1860 work on the kinetic theory.

\section{Proposition IV -- physical justification of its assumptions}\label{sect_justification}

\subsection{(A1) The existence of a stationary velocity distribution}\label{subsect_existence}

Let us now return to Maxwell's Proposition IV. The paragraph preceding the statement of Proposition IV, which we have already quoted, claims that after large number of collisions the kinetic energy ``will be divided among the particles according to some regular law, the average number of particles whose velocity lies between certain limits being ascertainable''  \citep[\pagenum 289]{Maxwell1860a}, meaning that this number is not going to be further altered by subsequent collisions, that is, that the velocity distribution becomes stationary. 

Indeed without the assumption of stationarity it would be hard to understand what enabled the presumption of the time-independent velocity distribution $f$ right in the beginning of the proof of Proposition IV. But if Maxwell succeeded to show that particle collisions lead to stationarity prior to Proposition IV then it would have been sufficient for him to assume, for the purposes of the proof of Proposition IV, the existence of the right kind of particle collisions. And this is what Maxwell actually says he does, to wit:
\begin{quote}
Prop. IV. To find the average number of particles whose velocities lie between given limits, {\em after a great number of collisions among a great number of equal particles.} (Emphasis added.)
\end{quote}
That his analysis of particle collisions was in part intended to establish the existence of a stationary velocity distribution is also clear from Maxwell's later remarks about his 1860 proof:
\begin{quote}
The only case in which I have determined the form of this [velocity distribution] function [in the 1860 paper] is that of one or more kinds of {\em molecules} which have {\em by their continual encounters brought about a distribution of velocity} such {\em that} the number of molecules whose velocity lies within given limits {\em remains constant}. \citep[\pagenum 437]{Maxwell1867} (Emphasis added.)
\end{quote}

In addition we have also seen that the idea that molecular collisions could lead to a stationary velocity distribution was more than baseless wishful thinking. We showed that Propositions I--III are indeed sufficient for a proof of tendency towards equilibrium, which is what Maxwell intended to do with his Proposition VI. In the special case when the masses are equal, which is the relevant case for Proposition IV, we have also seen that the proof of equalization of kinetic energies becomes so simple as to potentially warrant only verbal referencing. The availability of a proof of reaching equalization of temperatures lends credence for the availability of a proof of reaching a stationary velocity distributions.

Thus we can conclude that Maxwell's argument consists of two steps: a `lemma' showing that particle collisions lead to stationarity, and a `theorem' that makes use of the lemma. It is then of course no surprise that reference to particle collisions only appears in the statement of the theorem and is absent from the proof following the statement, since all the work having to do with particle collisions was previously delegated to the lemma; the proof of the theorem itself only makes use of the assumption of stationarity that follows from the lemma and its assumption about collisions. Establishing the conclusion of course requires both establishing the lemma and establishing the theorem, and thus Maxwell's 1860 argument does crucially depend on particle collisions, contrary to what historians of science (i.e. \citet[\pagenum 8]{Brush-Everitt-Garber1986b}, \citet[\pagenum 948]{Uffink2007}), who only focus on the proof that follows the statement of Proposition IV, claim.

The extent to which Maxwell succeeded in establishing the lemma is, of course, another question. Establishing that a certain tendency towards thermal equilibrium does follow from the model assumptions is clearly significant, but it is unclear to me how equalization of kinetic energies would strictly imply the onset of a stationary velocity distribution. If we conceptually divide the particles of the gas into two large groups -- say, red and blue -- then Maxwell's results imply that after collisions continue for a long time the mean velocity of all red particles will equal the mean velocity of all blue particles. This puts a constraint on how the velocity distribution of all particles may change after collisions. Since the division of particles to red and blue is arbitrary we end up with many such similar constraints, and jointly all these constraints {\em may} be sufficient to force the velocity distribution to become stationary. So much is clear, but whether this is so requires an explicit argument rather than a hopeful gesture.

Although this attempt stays close to the line of reasoning Maxwell himself utilizes in the end of his proof of Proposition VI he does not elaborate on the details of such or similar musings. He apparently thought that the evidence provided by Propositions I--III on the nature of particle collisions and his scattered remarks preceding the statement of Proposition IV are sufficient to establish that collisions lead to a stationary velocity distribution straightforwardly. Some followers of Maxwell shared this view, although it progressively became more clear that a proof of it, if possible at all, may not be easily befitted to a margin. Yet as late as 1890 we find proofs to the effect that Propositions I--III and Condition A implies stationarity:
\begin{quote}
[Proposition] 6. (a) Every distribution of velocities among the molecules which satisfies the condition that for given [velocity of the center of mass of a pair of colliding particles] $V$ all directions of [the velocity of rebound relative to the center of mass] $R$ are equally probable is [...] stationary. \citep[\pagenum 299]{Burbury1890}
\end{quote}
One of the main difficulties with these proofs is that they sneak in a probabilistic independence assumption in one form or another without appropriate justification.

\subsection{(A2) Probabilistic independence of velocity components}\label{subsect_independence}

\begin{figure}
	\begin{center}
		\includegraphics[height=12cm]{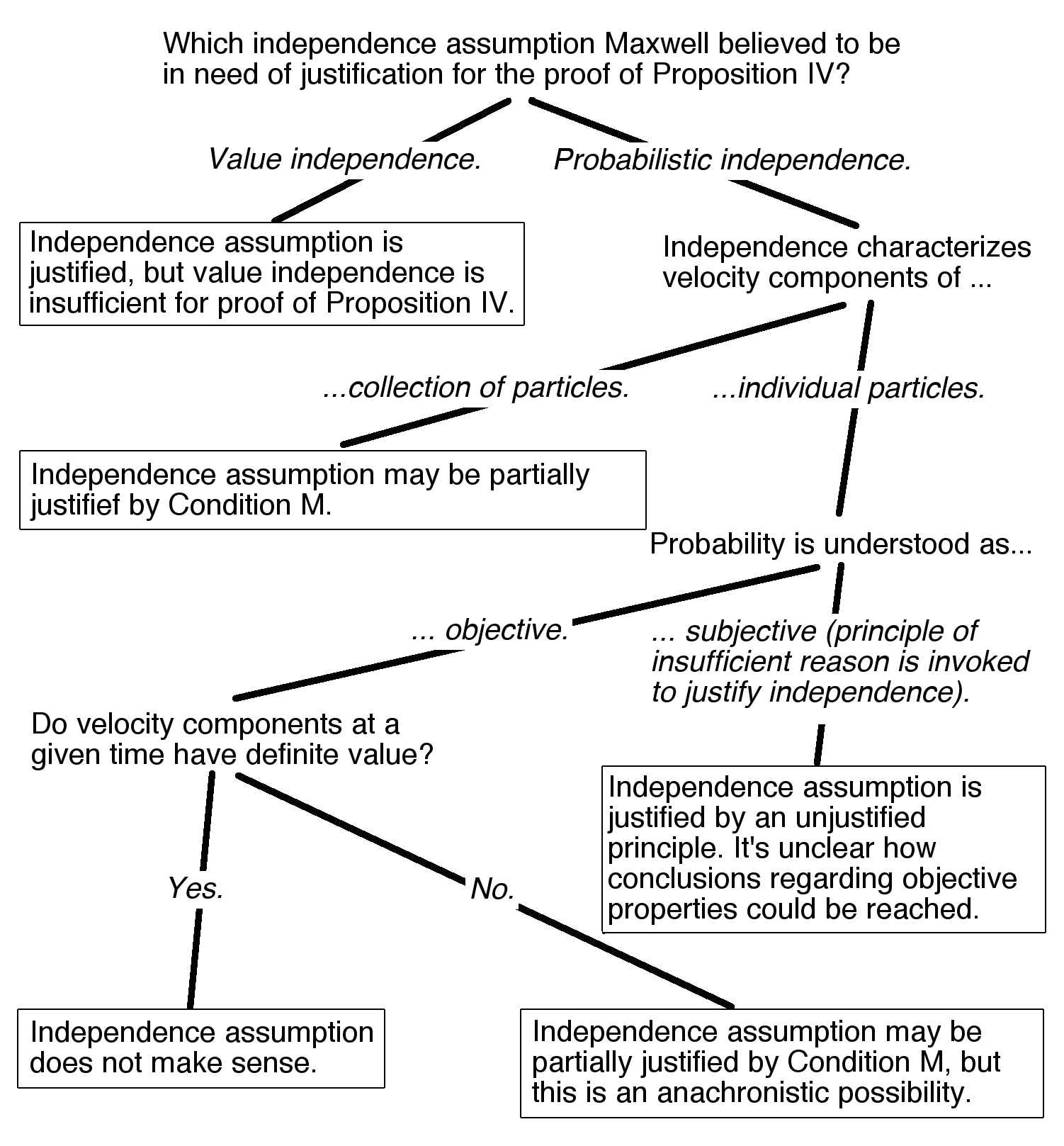}
		\caption{Various ways of understanding the term ``independent'' in the proof of Proposition IV.}\label{flowchartfigure}
	\end{center}
\end{figure}

The most crucial assumption of the proof of Proposition IV is the factorization of the velocity distribution to its marginals, which is a way to state the probabilistic independence of the velocity components. Let us re-quote, in its entirety, what Maxwell had to say about establishing this latter:
\begin{quote}
[...] the existence of the velocity $v_x$ does not in any way affect that of the velocities $v_y$ or $v_z$, since these are all at right angles to each other and independent[.]
\end{quote}
The obvious reading of this claim is that setting any value for the velocity component $v_x$ does not affect what values $v_y$ and $v_z$ could possibly take; in other words, that $v_x$ is value independent from $v_y$ and $v_z$. This obvious reading renders the claim obviously true and in no need of further justification. As we have seen before, Herschel emphasizes that value independence is the crucial assumption that is needed to justify the factorization of the distribution to its marginals. Given that Maxwell himself consistently used the term ``independent'' in the sense of value independence in his other works prior to 1860 it is then possible that he simply took what Herschel said for granted, and only bothered to point out that value independence of the velocity components in different directions clearly holds. Value independence of course does not imply probabilistic independence, and hence the assumption that would have been truly required for the mathematical proof remained unjustified.

According to another reading Maxwell was merely sloppy stating his assumption in the quoted claim, and what he really meant was that setting any value for the velocity component $v_x$ does not affect the {\em probability} of $v_y$ and $v_z$ taking their values. This reading seemingly finds support in what he says in his 1867 recollection that the 1860 proof was
\begin{quote}
[...] founded on the assumption that the probability of a molecule having a velocity resolved parallel to $x$ lying between given limits is not in any way affected by the knowledge that the molecule has a given velocity resolved parallel to $y$. \citep[\pagenum 437]{Maxwell1867}
\end{quote}
Aside from the curious allusion to `knowledge' this recollection correctly states that the proof of Proposition IV requires the assumption of probabilistic independence of $v_x$ and $v_y$. What the recollection does not make clear, however, is whether Maxwell also intended to rely on the assumption of probabilistic independence (as opposed to the assumption of value independence) {\em back in 1860}, and so this later reflection of Maxwell in itself can not be taken as evidence for an awareness of a distinction between probabilistic independence and value independence at the time when he produced his first proof. All we know is that in 1867 Maxwell found this reformulated independence requirement sufficiently ``precarious'' to warrant the construction of a different proof.

In the rest of this section we follow the latter reading, namely that Maxwell intended to have assumed a sort of probabilistic independence. It is still not entirely clear what could have been the intended carrier of the assumption of probabilistic independence according to Maxwell's 1867 recollection (see Figure \ref{flowchartfigure} for some possibilities). Although Maxwell seemingly talks about the velocity components of an individual particle, the velocity components of individual particles have a definite value at any given time, rendering any talk about a particle having a velocity component with some probability obscure (unless we either interpret probability as ignorance -- the above mentioned appearance of the term `knowledge' may indicate that -- or unless with a great dose of anachronism we give up definiteness of velocity values along the lines of an objective interpretation of the collision probabilities). It is more in line with the frequentist allures of the 1860 proof to take the probabilistic independence assumption to be a statement about a collection of particles, namely that the distribution of velocities along the $y$ axis for all particles is the same as the distribution of velocities along the $y$ axis for the particles whose velocity component along the $x$ axis is set to some value.

Does the {\em Illustrations} provide a justification for the assumption of probabilistic independence of velocity components of a collection of particles, or does at least have the resources to do so? With our reconstruction of Propositions I--III at hand the obvious guess is to trace back the assumption of probabilistic independence to Condition M. Recall that Condition M states that whenever two particles collide the impact parameter $\rho$ is distributed uniformly within a circle $C$. The `whenever' part of this assumption presumably states that $\rho$ is distributed uniformly regardless what are the incoming velocities of the pair of colliding particles or what is the configuration of other particles that are not participating in the collision. In a precise formulation this would presumably translate to conditional probabilistic independence of $\rho$ from the incoming velocities of the pair and from other physical quantities characterizing the remaining particles. If so then Condition M already furnishes a sort of probabilistic independence assumption regarding the velocities of a particle pair, and it is not inconceivable that some clever aggregation procedure would then lend credence to the assumption of probabilistic independence of velocity components of a collection of particles.

Maxwell himself does not hint at such argument in the {\em Illustrations}, but associating Condition M (and its consequence, Condition A) with a sense of probabilistic independence that is required for a derivation of the velocity distribution was so deep that even in 1894, during an ignited debate on the pages of {\em Nature} about the validity of Boltzmann's H-theorem, Condition A and the Stosszahlansatz (the crucial independence assumption about factorization of the velocity probability distribution that appears in Maxwell's 1867 proof) gets identified and interchanged without justification or warning (for a detailed account see \citet{Dias1994}). If the non-interchangeability of these assumptions did not occur to prominent physicists who had the advantage of many years of developments that followed Maxwell's 1860 proof, it seems reasonable to assume that it might have not occurred to Maxwell either. Hence Maxwell might have thought that Condition M, whose geometrical meaning is at least apparent, and which (as \citet{Gyenis2005} ventures) may be loosely motivated on the basis of Clausius' mean free path approach, justifies the probabilistic independence of $v_x$ and $v_y$.

It would be interesting to see, but I do not know of a formally satisfying investigation of the relationship of Condition M and the assumption of probabilistic independence of velocity components of a collection of particles. Clearly, any such investigation would require that first someone says exactly what Condition M states.

\subsection{(A3) Rotational symmetry of the velocity distribution}\label{subsect_rotationalsymmetry}

Maxwell's justification of the assumption of rotational symmetry of the velocity distribution is physically intuitive: nothing in the argument he gave for the factorization of the velocity distribution depends on the particular choice of the oblique coordinate system, and the physical behavior of the colliding particles should clearly not depend on how we choose to describe their behavior. Requiring that the factorization holds for other oblique coordinate systems entails that the distribution only depends from the distance from the origin. This result also follows when we only resort to non-dependence from choice of orthogonal coordinate systems, in which case it can be conveniently mathematically expressed as
		$$f(v_x) f(v_y) f(v_z) = \phi (v_x^2 + v_y^2 + v_z^2),$$
which is what Maxwell writes. (Note that the orthogonality assumption is not essential, only convenient. With any oblique coordinate system the functional equation (\ref{functionalequation}) still has a unique solution, a generalized normal distribution.)

To sum up, among the three mathematical assumptions of Maxwell's Proposition IV (A1) stationarity was intended to be justified on the basis of prior assumptions about particle collisions, (A2) probabilistic independence of velocity components was either misunderstood as value independence or was intended to be justified on the basis of prior assumptions about particle collisions, and (A3) rotational symmetry was supposed to follow from the non-dependence of the argument for the assumption of probabilistic independence of velocity components from the choice of coordinate system. Maxwell's analysis and assumptions regarding particle collisions hence played a crucial motivational role for the mathematical assumptions of Proposition IV.

Finally note that if the number of particles is finite then the actual distribution of velocities at any time must be discrete and have a bounded support. However the only measurable solution of the functional equation (\ref{functionalequation}), the Gaussian (\ref{gaussian}), is a continuous function with infinite support. To reconcile this tension one would need to give up at least one of the assumptions: one could insist that stationarity, rotational symmetry, or probabilistic independence only holds approximately, or that Maxwell's velocity distribution $f$ merely provides an approximate description. This is one of a number of issues relating to Proposition IV that we leave here undiscussed as they have been thoroughly addressed in the literature; for an overview see \citet[\ppagenum 945--946]{Uffink2007}.

\section{Conclusions}

The mathematical essence of the proof of Proposition IV is that a velocity distribution $\mathbf{f}(\vec{v})$ follows the Maxwell velocity distribution law if it satisfies the mathematical assumptions of probabilistic independence $\mathbf{f}(\vec{v}) = \mathbf{f}_x(v_x) \mathbf{f}_y(v_y) \mathbf{f}_z(v_z)$ and rotational symmetry $\mathbf{f}(\vec{v}) = f(|\vec{v}|)$. The mathematical essence of a proof may however mislead about what in it needs to be endowed with physical meaning and justification. We argued that emphasizing the importance of the implicit assumption of the existence of a stationary velocity distribution $\mathbf{f}(\vec{v})$ make the need for an argument for an approach to equilibrium clear and this in turn may shed light on the role Maxwell's widely ignored molecular collision model played in his {\em Illustrations of the Dynamical Theory of Gases}. In particular we showed that for perfectly elastic particles Maxwell's Proposition I--III -- under the assumptions that their time evolution satisfies the constraint that it can be split into successive time intervals during which each particle has an approximately equal independent chance to participate in at most one, pairwise collision, and that in this interval the impact parameter is approximately equally distributed on the collision circle (Condition M) -- yields a good ``bad'' mechanical proof of approach to equilibrium temperature, a precursor to Boltzmann's H-theorem, which also lends partial credence to the availability of a proof of reaching a stationary velocity distribution. We argued that Condition M of Proposition II furnishes a sort of probabilistic independence assumption regarding velocities that could have motivated the assumption of probabilistic independence, but we also suggested another potential explanation for the apparent lack of effort on the part of Maxwell to justify probabilistic independence, namely that Maxwell might have conflated the notions of probabilistic independence and value independence. We argued that distinction of meaning of these two notions was not yet available in the 1850's and we presented circumstantial evidence for these claims on the basis of relevant prior works of Clausius, Herschel, and Maxwell.

Section \ref{sect_justification} explored the ways how Propositions I--III, Condition M, and the approach to equilibrium furnished by a reconstructed version of Proposition VI could have played a role in Maxwell's physical justification of the three mathematical assumptions of Proposition IV; we have seen that there are multiple interpretative possibilities. In closing let me choose the interpretation which seems to me most plausible. On a physical basis Maxwell knew that after a sufficient amount of time a box of gas settles into a thermal equilibrium, and he made the important realization that supplementing a standard Newtonian collision model with the seemingly innocuous Condition M allows the proof of such an approach to thermal equilibrium if the particles are modeled as perfectly elastic spheres. Maxwell implicitly assumed that the onset of this equilibrium also entails that the velocity distribution function becomes stationary. Thinking that the existence of a stationary velocity distribution can be proved on a physical basis he looked around for mathematical proofs that could be used to characterize properties of such stationary distributions, and he recalled Herschel's characterization of the distribution of errors from reading his {\em Essays} as such. Unfortunately Herschel mischaracterized the crucial independence assumption of his own proof, and Maxwell implemented Herschel's proof together with this mischaracterization to the context of the kinetic theory. In turn the physical justification of the so-mischaracterized independence assumption (and as a consequence also the assumption of rotational symmetry) seemed straightforward. Maxwell later realized that this independence assumption is ``precarious'' and since other considerations also led him to abandon Clausius' mean free path approach that lies behind the collision model, he also abandoned the attempt to establish stationarity of the velocity distribution via Condition M. It took several decades until the scientific community came to a clear grasp of the distinctions between various interpretations of probability and various notions of independence, and thus Maxwell's early mistake in his first paper on the kinetic theory was by no means an obvious one.

\section*{Acknowledgement}

This paper is based on a substantive revision of the (unpublished) work \citet{Gyenis2005}. I thank John Norton, John Earman, Jos Uffink, Michael Strevens, Michael Reed, and M\'arton G\"om\"ori for feedback. (I also thank Daniel Marg\'ocsy and audiences of the Philosophy of Science Reading Group of the Institute of Philosophy, HAS RCH and of the Theoretical Philosophy Forum, ELTE for feedback on briefly mentioned parts of this work that got removed at the request of a reviewer.) 

This work was supported by the National Research, Development and Innovation Office, K 115593 grant.

\newpage

\section*{Appendix}

Let $\vec{v}$ and $\vec{V}$ be the incoming velocities and $\vec{v}\V$ and $\vec{V}\V$ be the rebound velocities of two spheres of masses $m$ and $M$ that undergo a perfectly elastic collision. Due to conservation of momentum the velocity of the center of mass is unchanged by the collision:
\begin{equation}
	\vec{v}_{CM} \doteq \frac{m \vec{v} + M \vec{V}}{m+M} = \frac{m \vec{v}\V + M \vec{V}\V}{m+M} \doteq \vec{v}\V_{CM}.
\end{equation}

Let $\vec{r}$, $\vec{R}$ (and $\vec{r}\V$, $\vec{R}\V$) denote the incoming (and rebound) velocities relative to this center of mass, that is
\begin{eqnarray}
	\vec{r} 	& \doteq & \vec{v}-\vec{v}_{CM} = \vec{v} - \frac{m \vec{v} + M \vec{V}}{m+M} = \frac{M}{m+M} (\vec{v}-\vec{V}) \\
	\vec{R} 	& \doteq & \vec{V}-\vec{v}_{CM} = \vec{V} - \frac{m \vec{v} + M \vec{V}}{m+M} = \frac{m}{m+M} (\vec{V}-\vec{v}) \\
	\vec{r}\V 	& \doteq & \vec{v}\V-\vec{v}\V_{CM} = \vec{v}\V-\vec{v}_{CM} \label{eq_appendix_4} \\
	\vec{R}\V 	& \doteq & \vec{V}\V-\vec{v}\V_{CM} = \vec{V}\V-\vec{v}_{CM}, \label{eq_appendix_5}
\end{eqnarray}
and hence noting that $v \doteq \| \vec{v} \|$, $V \doteq \| \vec{V} \|$, $v' \doteq \| \vec{v}\V \|$ etc. we get
\begin{eqnarray}
	r^2   		& = & \vec{r} \cdot \vec{r} = \left(\frac{M}{m+M}\right)^2 \left( \vec{v}\cdot \vec{v} + \vec{V}\cdot \vec{V} - 2 \vec{v} \cdot \vec{V} \right) \\
				& = & \left( \frac{M}{m+M} \right)^2 \left( v^2 + V^2 - 2 v V \cos(\alpha) \right), \label{eq_appendix_7} \\
	R^2			& = & \vec{R} \cdot \vec{R} = \left(\frac{m}{m+M}\right)^2 \left( \vec{v}\cdot \vec{v} + \vec{V}\cdot \vec{V} - 2 \vec{v} \cdot \vec{V} \right) \\
				& = & \left(\frac{m}{m+M}\right)^2 \left( v^2 + V^2 - 2 v V \cos(\alpha) \right),  \label{eq_appendix_9} \\
	v_{CM}^2	& = & \vec{v}_{CM} \cdot \vec{v}_{CM} = \frac{m \vec{v} + M \vec{V}}{m+M} \cdot \frac{m \vec{v} + M \vec{V}}{m+M} =  \\
				& = & \left(\frac{1}{m+M}\right)^2 \left( m^2 \vec{v}\cdot \vec{v} + M^2 \vec{V}\cdot \vec{V} + 2 m M \vec{v} \cdot \vec{V} \right) \\
				& = & \left(\frac{1}{m+M}\right)^2 \left( m^2 v^2 + M^2 V^2 + 2mMvV \cos(\alpha) \right)  \label{eq_appendix_12}
\end{eqnarray}
with $\alpha \doteq \angle (\vec{v}, \vec{V})$. 

From (\ref{eq_appendix_4}), (\ref{eq_appendix_5}), and from $r' = r$, $R' = R$ (which we get from (\ref{eq_r_is_unchanging}) where we utilized that the collision is perfectly elastic) we get
\begin{eqnarray}
	v'^2	& = & (\vec{v}_{CM} + \vec{r}\V ) \cdot (\vec{v}_{CM} + \vec{r}\V) = v_{CM}^2 + r'^2 + 2 \vec{v}_{CM} \cdot \vec{r}\V \\
			& = & v_{CM}^2 + r^2 + 2 v_{CM} r \cos(\gamma) \label{eq_appendix_14} \\
	V'^2	& = & (\vec{v}_{CM} + \vec{R}\V ) \cdot (\vec{v}_{CM} + \vec{R}\V) = v_{CM}^2 + R'^2 + 2 \vec{v}_{CM} \cdot \vec{R}\V \\
			& = & v_{CM}^2 + R^2 - 2 v_{CM} R \cos(\gamma) \label{eq_appendix_16}
\end{eqnarray}
with $\gamma \doteq \angle (\vec{v}_{CM}, \vec{r}\V)$ and noting that $\angle (\vec{v}_{CM}, \vec{R}\V) = 180^o - \angle (\vec{v}_{CM}, \vec{r}\V) $.

Multiplying (\ref{eq_appendix_14}) with $m/2$, (\ref{eq_appendix_16}) with $M/2$, subtracting them from each other, and making judicious substitutions for $r^2$  (eq. (\ref{eq_appendix_7})), $R^2$  (eq. (\ref{eq_appendix_9})), and $v_{CM}^2$  (eq. (\ref{eq_appendix_12})) yields
\begin{eqnarray*}
	\frac{m v'^2}{2} - \frac{M V'^2}{2} 	& = & \frac{m}{2} \left( v_{CM}^2 + r^2 + 2 v_{CM} r \cos(\gamma) \right) \\
											&	& - \frac{M}{2} \left( v_{CM}^2 + R^2 - 2 v_{CM} R \cos(\gamma) \right) \\
											& = & \frac{m}{2} \left(\frac{1}{m+M}\right)^2 \left( m^2 v^2 + M^2 V^2 + 2mMvV \cos(\alpha) \right) \\
											&	& + \frac{m}{2} \left( \frac{M}{m+M} \right)^2 \left( v^2 + V^2 - 2 v V \cos(\alpha) \right) + \frac{m}{2} 2 v_{CM} r \cos(\gamma) \\
											&	& - \frac{M}{2} \left(\frac{1}{m+M}\right)^2 \left( m^2 v^2 + M^2 V^2 + 2mMvV \cos(\alpha) \right) \\
											&	& - \frac{M}{2} \left(\frac{m}{m+M}\right)^2 \left( v^2 + V^2 - 2 v V \cos(\alpha) \right) + \frac{M}{2} 2 v_{CM} R \cos(\gamma) \\
											& = & \frac{1}{2} \left(\frac{1}{m+M} \right)^2 ( m^3 v^2 + m M^2 V^2 + 2 m^2 M v V cos(\alpha) \\
											&	& ~~~~~~~~~~~~~~~~~~~ + m M^2 v^2 + m M^2 V^2 - 2 m M^2 v V \cos(\alpha)  \\
											&	& ~~~~~~~~~~~~~~~~~~~ - m^2 M v^2 - M^3 V^2 - 2 m M^2 v V \cos(\alpha) \\
											&	& ~~~~~~~~~~~~~~~~~~~ - m^2 M v^2 - m^2 M V^2 + 2 m^2 M v V \cos(\alpha) ) \\
											&	& + (m r + M R) v_{CM} \cos(\gamma) \\
											& = & + \frac{1}{2} \left(\frac{1}{m+M} \right)^2 ( m v^2 (m^2 + M^2 - 2 m M) \\
											&	& ~~~~~~~~~~~~~~~~~~~ - M V^2 (m^2 + M^2 - 2 m M)  ) \\
											&   & + 2 \left(\frac{1}{m+M} \right)^2 (m-M) mMvV\cos(\alpha) \\
											&	& + (m r + M R) v_{CM} \cos(\gamma) \\
											& = & C_0 \cdot \left(\frac{m v^2}{2} - \frac{M V^2}{2} \right) + C_1 \cdot \cos(\alpha) + C_2 \cdot \cos(\gamma)
\end{eqnarray*}
with
\begin{eqnarray}
	C_0 & \doteq &	\left(\frac{m-M}{m+M}\right)^2, \\
	C_1 & \doteq &	2 \frac{(m - M)}{{(m + M)^2}} m M v V, \\
	C_2 & \doteq & (m r + M R) v_{CM},
\end{eqnarray}
which proves equation (\ref{eq_differenceofkineticenergiesafterrebound}). Clearly, $0 \leq C_0 < 1$.

\section*{References}


\begin{thebibliography}{32}
\expandafter\ifx\csname natexlab\endcsname\relax\def\natexlab#1{#1}\fi
\providecommand{\url}[1]{\texttt{#1}}
\providecommand{\href}[2]{#2}
\providecommand{\path}[1]{#1}
\providecommand{\DOIprefix}{doi:}
\providecommand{\ArXivprefix}{arXiv:}
\providecommand{\URLprefix}{URL: }
\providecommand{\Pubmedprefix}{pmid:}
\providecommand{\doi}[1]{\href{http://dx.doi.org/#1}{\path{#1}}}
\providecommand{\Pubmed}[1]{\href{pmid:#1}{\path{#1}}}
\providecommand{\bibinfo}[2]{#2}
\ifx\xfnm\relax \def\xfnm[#1]{\unskip,\space#1}\fi
\bibitem[{Boltzmann(1868)}]{Boltzmann1868}
\bibinfo{author}{Boltzmann, L.} (\bibinfo{year}{1868}).
\newblock \bibinfo{title}{Studien \"uber das Gleichgewicht der lebendigen Kraft zwischen bewegten}.
\newblock {\it \bibinfo{journal}{Wiener Berichte}\/},  {\it
  \bibinfo{volume}{58}\/}, \bibinfo{pages}{517--560}.
\bibitem[{Brush(1958)}]{Brush1958}
\bibinfo{author}{Brush, S.~G.} (\bibinfo{year}{1958}).
\newblock \bibinfo{title}{The Development of Kinetic Theory of Gases: IV Maxwell}.
\newblock {\it \bibinfo{journal}{Annals of Science}\/},  {\it
  \bibinfo{volume}{14}\/}, \bibinfo{pages}{243--255}.
\bibitem[{Brush(1971)}]{Brush1971}
\bibinfo{author}{Brush, S.~G.} (\bibinfo{year}{1971}).
\newblock \bibinfo{title}{James Clerk Maxwell and the kinetic theory of gases:
  A review based on recent historical studies}.
\newblock {\it \bibinfo{journal}{American Journal of Physics}\/},  {\it
  \bibinfo{volume}{31}\/}, \bibinfo{pages}{631--640}.
\bibitem[{Brush(1976)}]{Brush1976}
\bibinfo{author}{Brush, S.~G.} (\bibinfo{year}{1976}).
\newblock {\it \bibinfo{title}{The Kind of Motion we call Heat}\/}
  volume~\bibinfo{volume}{VI} of {\it \bibinfo{series}{Studies in Statistical
  Mechanics}\/}.
\newblock \bibinfo{address}{Amsterdam, New York, Oxford}:
  \bibinfo{publisher}{North-Holland Publishing Company}.
\bibitem[{Brush(1983)}]{Brush1983}
\bibinfo{author}{Brush, S.~G.} (\bibinfo{year}{1983}).
\newblock {\it \bibinfo{title}{Statistical Physics and the Atomic Theory of
  Matter From Boyle and Newton to Landau and Onsager}\/}.
\newblock Princeton Series in Physics.
\newblock \bibinfo{address}{Princeton, New Jersey}:
  \bibinfo{publisher}{Princeton University Pres}.
\bibitem[{Brush(2003)}]{Brush2003}
\bibinfo{author}{Brush, S.~G.} (\bibinfo{year}{2003}).
\newblock {\it \bibinfo{title}{The Kinetic Theory of Gases: Anthology of
  Historical Papers with Historical Commentary}\/}.
\newblock \bibinfo{address}{London}: \bibinfo{publisher}{Imperial College
  Press}.
\bibitem[{Brush et~al.(1986{\natexlab{a}})Brush, Everitt \&
  Garber}]{Brush-Everitt-Garber1986b}
\bibinfo{author}{Brush, S.~G.}, \bibinfo{author}{Everitt, C. W.~F.}, \&
  \bibinfo{author}{Garber, E.} (\bibinfo{year}{1986}{\natexlab{a}}).
\newblock \bibinfo{title}{Kinetic theory and the properties of gases: Maxwell's
  work in its nineteenth-century context}.
\newblock In  \cite{Brush-Everitt-Garber1986a}.
\bibitem[{Brush et~al.(1986{\natexlab{b}})Brush, Everitt \&
  Garber}]{Brush-Everitt-Garber1986a}
\bibinfo{editor}{Brush, S.~G.}, \bibinfo{editor}{Everitt, C. W.~F.}, \&
  \bibinfo{editor}{Garber, E.} (Eds.) (\bibinfo{year}{1986}{\natexlab{b}}).
\newblock {\it \bibinfo{title}{Maxwell on Molecules and Gases}\/}.
\newblock \bibinfo{address}{Cambridge, Massachusetts}: \bibinfo{publisher}{The
  MIT Press}.
\bibitem[{Burbury(1890)}]{Burbury1890}
\bibinfo{author}{Burbury, S.~H.} (\bibinfo{year}{1890}).
\newblock \bibinfo{title}{On some problems in the kinetic theory of gases}.
\newblock {\it \bibinfo{journal}{Philosophical Magazine}\/},  {\it
  \bibinfo{volume}{30}\/}, \bibinfo{pages}{298--317}.
\bibitem[{Burbury(1894)}]{Burbury1894a}
\bibinfo{author}{Burbury, S.~H.} (\bibinfo{year}{1894}).
\newblock \bibinfo{title}{Boltzmann's minimum function}.
\newblock {\it \bibinfo{journal}{Nature}\/},  {\it \bibinfo{volume}{51}\/},
  \bibinfo{pages}{78--79}.
\bibitem[{Buys-Ballot(1858)}]{Buys-Ballot1858}
\bibinfo{author}{Buys-Ballot, C. H.~D.} (\bibinfo{year}{1858}).
\newblock \bibinfo{title}{Ueber die Art von Bewegung, welche wir W\"arme und
  Elektrizit\"at nennen}.
\newblock {\it \bibinfo{journal}{Annalen der Physik}\/},  {\it
  \bibinfo{volume}{103}\/}, \bibinfo{pages}{240--259}.
\bibitem[{Campbell \& Garnett(1882)}]{Campbell-Garnett1882}
\bibinfo{author}{Campbell, L.}, \& \bibinfo{author}{Garnett, W.}
  (\bibinfo{year}{1882}).
\newblock {\it \bibinfo{title}{The Life of James Clerk Maxwell}\/}.
\newblock \bibinfo{address}{London}: \bibinfo{publisher}{Macmillan and Co.}
\bibitem[{Clausius(1857{\natexlab{a}})}]{Clausius1857}
\bibinfo{author}{Clausius, R.} (\bibinfo{year}{1857}{\natexlab{a}}).
\newblock \bibinfo{title}{The nature of motion which we call heat}.
\newblock {\it \bibinfo{journal}{Philosophical Magazine}\/},  {\it
  \bibinfo{volume}{14}\/}, \bibinfo{pages}{108--127}.
\newblock \bibinfo{note}{English translation of his \citeyear{Clausius1857de}.
  See also in \citet{Brush2003}, 111--134.}
\bibitem[{Clausius(1857{\natexlab{b}})}]{Clausius1857de}
\bibinfo{author}{Clausius, R.} (\bibinfo{year}{1857}{\natexlab{b}}).
\newblock \bibinfo{title}{Ueber die Art der Bewegung, welche wir W\"arme nennen}.
\newblock {\it \bibinfo{journal}{Annalen der Physik}\/},  {\it
  \bibinfo{volume}{100}\/}, \bibinfo{pages}{353--380}.
\bibitem[{Clausius(1858)}]{Clausius1858}
\bibinfo{author}{Clausius, R.} (\bibinfo{year}{1858}).
\newblock \bibinfo{title}{Ueber die Mittlere L\"ange der Were, welche bei Molecularbewegung gasf\"ormigen K\"orper von den einzelnen Molec\"ulen zur\"uckgelegt werden, nebst einigen anderen Bemerkungen \"uber die mechanischen W\"armetheorie}.
\newblock {\it \bibinfo{journal}{Annalen der Physik}\/},  {\it
  \bibinfo{volume}{105}\/}, \bibinfo{pages}{239--258}.
\bibitem[{Clausius(1859)}]{Clausius1859}
\bibinfo{author}{Clausius, R.} (\bibinfo{year}{1859}).
\newblock \bibinfo{title}{On the mean lengths of the paths described by the
  separate molecules of gaseous bodies}.
\newblock {\it \bibinfo{journal}{Philosophical Magazine}\/},  {\it
  \bibinfo{volume}{17}\/}, \bibinfo{pages}{81--01}.
\newblock \bibinfo{note}{English translation of his \citeyear{Clausius1858}.
  See also in \citet{Brush2003}, 135--147.}
\bibitem[{Clausius(1862)}]{Clausius1862}
\bibinfo{author}{Clausius, R.} (\bibinfo{year}{1862}).
\newblock \bibinfo{title}{On the conduction of heat by gases}.
\newblock {\it \bibinfo{journal}{Philosophical Magazine}\/},  {\it
  \bibinfo{volume}{23}\/}, \bibinfo{pages}{416--435, 512--535}.
\bibitem[{Dias(1994)}]{Dias1994}
\bibinfo{author}{Dias, P. M.~C.} (\bibinfo{year}{1994}).
\newblock \bibinfo{title}{``Will someone say exactly what the H-theorem
  proves?'' A study of Burbury's Condition A and Maxwell's Proposition II}.
\newblock {\it \bibinfo{journal}{Archive for History of Exact Sciences}\/},
  {\it \bibinfo{volume}{46}\/}, \bibinfo{pages}{341--366}.
\bibitem[{Garber(1970)}]{Garber1970}
\bibinfo{author}{Garber, E.~W.} (\bibinfo{year}{1970}).
\newblock \bibinfo{title}{Clausius and Maxwell's kinetic theory of gases}.
\newblock {\it \bibinfo{journal}{Hist. Studies Phys. Sci.}\/},  {\it
  \bibinfo{volume}{2}\/}, \bibinfo{pages}{299--319}.
\bibitem[{Gillispie(1963)}]{Gillispie1963}
\bibinfo{author}{Gillispie, C.~C.} (\bibinfo{year}{1963}).
\newblock \bibinfo{title}{Intellectual factors in the background of analysis by
  probabilities}.
\newblock In \bibinfo{editor}{A.~C. Crombie} (Ed.), {\it
  \bibinfo{booktitle}{Scientific Change}\/} (pp. \bibinfo{pages}{431--453}).
\newblock \bibinfo{address}{New York}: \bibinfo{publisher}{Basic Books}.
\bibitem[{Gyenis(2005)}]{Gyenis2005}
\bibinfo{author}{Gyenis, B.} (\bibinfo{year}{2005}).
\newblock \bibinfo{title}{Maxwell and the normal distribution}.
\newblock \bibinfo{note}{M.Sc. research paper, Department of History and
  Philosophy of Science, University of Pittsburgh}.
\bibitem[{Heimann(1970)}]{Heimann1970}
\bibinfo{author}{Heimann, P.~M.} (\bibinfo{year}{1970}).
\newblock \bibinfo{title}{Molecular forces, statistical representation and
  maxwell's demon}.
\newblock {\it \bibinfo{journal}{Hist. Phil. Sci.}\/},  {\it
  \bibinfo{volume}{1}\/}, \bibinfo{pages}{189--211}.
\bibitem[{Herschel(1850)}]{Herschel1850}
\bibinfo{author}{Herschel, J. F.~W.} (\bibinfo{year}{1850}).
\newblock \bibinfo{title}{Quetelet on probabilities}.
\newblock {\it \bibinfo{journal}{Edinburgh Rev.}\/},  {\it
  \bibinfo{volume}{92}\/}, \bibinfo{pages}{1--57}.
\bibitem[{Herschel(1857)}]{Herschel1857}
\bibinfo{author}{Herschel, J. F.~W.} (\bibinfo{year}{1857}).
\newblock {\it \bibinfo{title}{Essays from the ``Edinburgh" and ``Quarterly"
  Reviews}\/}.
\newblock \bibinfo{address}{London}: \bibinfo{publisher}{Layton}.
\bibitem[{Maxwell(1855)}]{Maxwell1855a}
\bibinfo{author}{Maxwell, J.~C.} (\bibinfo{year}{1855}).
\newblock \bibinfo{title}{On the theory of colours in relation to
  colour-blindness. a letter to dr G. Wilson}.
\newblock In \bibinfo{editor}{W.~D. Niven} (Ed.), {\it \bibinfo{booktitle}{The
  Scientific Papers of James Clerk Maxwell}\/} (pp. \bibinfo{pages}{119--125}).
\newblock \bibinfo{address}{New York}: \bibinfo{publisher}{Dover Publications,
  Inc.} volume~\bibinfo{volume}{I}. (\bibinfo{edition}{1965th} ed.).
\newblock \bibinfo{note}{Appeared in the {\it Transactions of the Royal Society
  of Arts}, Vol. IV. Part III.}
\bibitem[{Maxwell(1859)}]{Maxwell1859b}
\bibinfo{author}{Maxwell, J.~C.} (\bibinfo{year}{1859}).
\newblock \bibinfo{title}{Letter from Maxwell to George Gabriel Stokes, May 30,
  1859}.
\newblock In  \cite{Brush-Everitt-Garber1986a}.
\bibitem[{Maxwell(1860)}]{Maxwell1860a}
\bibinfo{author}{Maxwell, J.~C.} (\bibinfo{year}{1860}).
\newblock \bibinfo{title}{Illustrations of the dynamical theory of gases}.
\newblock {\it \bibinfo{journal}{Philosophical Magazine}\/},  {\it
  \bibinfo{volume}{19, 20}\/}, \bibinfo{pages}{19--32, 21--37}.
\newblock \bibinfo{note}{Quotations from \citet{Brush-Everitt-Garber1986a},
  286--318.}
\bibitem[{Maxwell(1863)}]{Maxwell1863}
\bibinfo{author}{Maxwell, J.~C.} (\bibinfo{year}{1863}).
\newblock \bibinfo{title}{On the conduction of heat in gases}.
\newblock In  \cite{Brush-Everitt-Garber1986a}.
\newblock \bibinfo{note}{Unpublished manuscript in which Maxwell tried to
  correct some of the omissions and errors of his 1860 paper.}
\bibitem[{Maxwell(1867)}]{Maxwell1867}
\bibinfo{author}{Maxwell, J.~C.} (\bibinfo{year}{1867}).
\newblock \bibinfo{title}{On the dynamical theory of gases}.
\newblock {\it \bibinfo{journal}{Proceedings of the Royal Society of
  London}\/},  {\it \bibinfo{volume}{15}\/}, \bibinfo{pages}{167--171}.
\newblock \bibinfo{note}{Also appeared in {\em Philosophical Magazine}, 32:
  129--145 (1866), 35: 129--145, 185--217 (1868). Quotations from
  \citet{Brush-Everitt-Garber1986a}, 420--472.}
\bibitem[{Porter(1981)}]{Porter1981}
\bibinfo{author}{Porter, T.} (\bibinfo{year}{1981}).
\newblock \bibinfo{title}{A statistical survey of gases: Maxwell's social
  physics}.
\newblock {\it \bibinfo{journal}{Historical Studies in the Physical
  Sciences}\/},  {\it \bibinfo{volume}{12}\/}, \bibinfo{pages}{77--116}.
\bibitem[{Truesdell(1975)}]{Truesdell1975}
\bibinfo{author}{Truesdell, C.} (\bibinfo{year}{1975}).
\newblock \bibinfo{title}{Early kinetic theories of gases}.
\newblock {\it \bibinfo{journal}{Arch. Hist. Exact Sci.}\/},  {\it
  \bibinfo{volume}{15}\/}, \bibinfo{pages}{1--66}.
\bibitem[{Uffink(2007)}]{Uffink2007}
\bibinfo{author}{Uffink, J.} (\bibinfo{year}{2007}).
\newblock \bibinfo{title}{Compendium of the foundations of classical
  statistical physics}.
\newblock In \bibinfo{editor}{J.~E. J.~Butterfield} (Ed.), {\it
  \bibinfo{booktitle}{Philosophy of Physics, Part B}\/} Handbook of the
  Philosophy of Science.
\newblock \bibinfo{address}{The Netherlands}: \bibinfo{publisher}{Elsevier}.

\end{thebibliography}

\end{document}